\documentclass[11pt]{article}
\usepackage{geometry}                
\usepackage{graphicx}
\usepackage{amssymb}
\usepackage{epstopdf}
\usepackage{mathtools}
\title{Equitability, mutual information, and the\\ maximal information coefficient}
\author{Justin B.\ Kinney\footnote{Please send correspondence to jkinney@cshl.edu} ~and Gurinder S.\ Atwal \\ 
\small{Simons Center for Quantitative Biology} \\ \small{Cold Spring Harbor Laboratory} \\ \small{Cold Spring Harbor, NY~11724}}
\date{}                                           

\newenvironment{definition}[1][Definition]{\begin{trivlist}
\item[\hskip \labelsep {\bfseries #1}]}{\end{trivlist}}

\begin{document}
\maketitle

\begin{abstract}
Reshef et al.\ recently proposed a new statistical measure, the ``maximal information coefficient'' (MIC), for quantifying arbitrary dependencies between pairs of stochastic quantities. MIC is based on mutual information, a fundamental quantity in information theory that is widely understood to serve this need. MIC, however, is not an estimate of mutual information. Indeed, it was claimed that MIC possesses a desirable mathematical property called ``equitability'' that mutual information lacks. This was not proven; instead it was argued solely through the analysis of simulated data. Here we show that this claim, in fact, is incorrect. First we offer mathematical proof that no (non-trivial) dependence measure satisfies the definition of equitability proposed by Reshef et al.. We then propose a self-consistent and more general definition of equitability that follows naturally from the Data Processing Inequality. Mutual information satisfies this new definition of equitability while MIC does not. Finally, we show that the simulation evidence offered by Reshef et al.\ was artifactual. We conclude that estimating mutual information is not only practical for many real-world applications, but also provides a natural solution to the problem of quantifying associations in large data sets. 
\end{abstract}


\section*{Introduction}

Reshef et al.\ \cite{Reshef:2011p834} discuss a basic problem in statistics. Given a large number of data points, each comprising a pair of real quantities $x$ and $y$, how can one reliably quantify the dependence between these two quantities without prior assumptions about the specific functional form of this dependence? For instance, Pearson  correlation can accurately quantify dependencies when the underlying relationship is linear and the noise is Gaussian, but typically provides poor quantification of noisy relationships that are non-monotonic. 

Reshef et al.\ argue that a good dependence measure should be ``equitable'' -- it ``should give similar scores to equally noisy relationships of different types'' \cite{Reshef:2011p834}. In other words, a measure of how much noise is in an $x$-$y$ scatter plot should not depend on what the specific functional relationship between $x$ and $y$ would be in the absence of noise. 

Soon after the inception of information theory \cite{Shannon:1949p1061}, a quantity now known as ``mutual information'' became recognized as providing a principled solution to the problem of quantifying arbitrary dependencies. If one has enough data to reconstruct the joint probability distribution $p(x,y)$, one can compute the corresponding mutual information $I[x;y]$. This quantity is zero if and only if $x$ and $y$ are independent; otherwise it has a value greater than zero, with larger values corresponding to stronger dependencies. These values have a fundamental meaning: $I[x;y]$ is the amount of information -- in units known as ``bits'' -- that the value of one variable reveals about the value of the other. Moreover, mutual information can be computed between any two types of random variables (real numbers, multidimensional vectors, qualitative categories, etc.), and $x$ and $y$ need not be of the same variable type. Mutual information also has a natural generalization, called multi-information, which quantifies dependencies between three or more variables \cite{Slonim:2005p433}. 

It has long been recognized that mutual information is able to quantify the strength of dependencies without regard to the specific functional form of those dependencies.\footnote{Indeed, $I[x;y]$ is invariant under arbitrary invertible transformations of $x$ or $y$ \cite{Cover:1991p1068}.} Reshef et al.\ claim, however, that mutual information does not satisfy their notion of equitability. Moreover, they claim that normalizing a specific estimate of mutual information into a new statistic they call the ``maximal information coefficient'' (MIC) is able to restore equitability. Both of these points were emphasized in a recent follow-up study \cite{Reshef:2013p1067}. However, neither the original study nor the follow-up provide any mathematical arguments for these claims. Instead, the authors argue this point solely by comparing estimates of mutual information and MIC on simulated data. 

Here we mathematically prove that these claims are wrong on a number of counts. After reviewing the definitions of mutual information and MIC, we prove that no non-trivial dependence measure, including MIC, satisfies the definition of equitability given by Reshef et al.. We then propose a new and more general definition of equitability, which we term ``self-equitability''. This definition takes the form of a simple self-consistency condition and is a direct consequence of the Data Processing Inequality (DPI) \cite{Cover:1991p1068}. Mutual information satisfies self-equitability while MIC does not. Simple examples demonstrating how MIC violates self-equitability and related notions of dependence are given. Finally, we revisit the simulations performed by Reshef et al., and find their evidence regarding the equitability of mutual information and MIC to be artifactual. Specifically, MIC appears equitable in both their original paper and their follow-up study because random fluctuations (caused by limited simulated data) obscure the systematic bias that results from this measure's non-equitability. Conversely, limited data combined with an inappropriate runtime parameter in the Kraskov et al.\ \cite{Kraskov:2004p909} estimation algorithm are responsible for the highly non-equitable behavior reported for mutual information. 

\section*{Mutual information}

Consider two real continuous stochastic variables $x$ and $y$, drawn from a joint probability distribution $p(x,y)$. The mutual information between these variables is defined as \cite{Cover:1991p1068},
\begin{eqnarray} 
I[x;y] = \int dx~dy~p(x,y) \log_2 \frac{p(x,y)}{p(x)p(y)}. \label{MI_def}
\end{eqnarray} 
where $p(x)$ and $p(y)$ are the marginal distributions of $p(x,y)$.\footnote{When $\log_2$ is used in Eq.\ \ref{MI_def}, mutual information is said to be given in units called ``bits''; if $\log_e$ is used instead, the units are referred to as ``nats''.} Defined as such, mutual information has a number of important properties. $I[x;y]$ is non-negative, with $I[x;y] = 0$ occurring only when $p(x,y) = p(x)p(y)$. Thus, mutual information will be greater than zero when $x$ and $y$ exhibit \emph{any} mutual dependence, regardless of how nonlinear that dependence is. Moreover, the stronger the mutual dependence, the larger the value of $I[x;y]$. In the limit where $y$ is a deterministic function of $x$, $I[x;y] = \infty$. 

Accurately estimating mutual information from finite data, however, is nontrivial. The difficulty lies in estimating the joint distribution $p(x,y)$ from a finite sample of $N$ data points. The simplest approach is to ``bin'' the data -- to superimpose a rectangular grid on the $x$-$y$ scatter plot, then assign each continuous $x$ value (or $y$ value) to the column bin $X$ (or row bin $Y$) into which it falls. Mutual information can then be estimated from the resulting discretized joint distribution $p(X,Y)$ as
\begin{eqnarray}
I[x;y] \approx I[X;Y] = \sum_{X,Y} p(X,Y) \log_2 \frac{p(X,Y)}{p(X)p(Y)},
\end{eqnarray}
where $p(X,Y)$ is the fraction of binned data falling into bin $(X,Y)$. Estimates of mutual information that rely on this simple binning procedure are called ``naive'' estimates. The problem with such naive estimates is that they systematically overestimate $I[x;y]$. This has long been recognized as a problem \cite{Treves:1995p244}, and significant attention has been devoted to providing other methods for accurately estimating mutual information (see Discussion). This estimation problem, however, becomes easier as $N$ becomes large. In the large data limit ($N \to \infty$), the joint distribution $p(x,y)$ can be determined to arbitrary accuracy, and thus so can $I[x;y]$. 

\section*{The maximal information coefficient}

Reshef et al.\ define MIC on a set of $N$ data points $(x,y)$ as follows,
\begin{eqnarray}
MIC[x;y] = \max_{|X||Y| < B} \frac{I[X;Y]}{\log_2(\min(|X|,|Y|))}. \label{MIC}
\end{eqnarray}
Specifically, one first adopts a binning scheme assigning each data point $(x,y)$ to a bin $(X,Y)$ as described above. The resulting naive mutual information estimate $I[X;Y]$ is then computed from the frequency table $p(X,Y)$. It is then divided by the log number of $X$ bins (denoted $|X|$) or $Y$ bins ($|Y|$), whichever is less. The resulting value will depend, of course, on the number of bins in both dimensions, as well as on where the bin boundaries are drawn. Reshef et al.\ define MIC as the quantity that results from maximizing this ratio over \emph{all} possible binning schemes for which the total number of bins, $|X||Y|$, is less than some number $B$.  The fact that $0 \le I[X;Y] \le \log_2( \min( |X|, |Y|))$ implies MIC will always fall between 0 and 1. 

At first this definition might seem like a sensible way to limit the number of bins one uses when discretizing data: if the resolution of the binning scheme is increased, the concomitant increase in $I[X;Y]$ must be enough to overcome the increase in the log number of bins. However, this normalization scheme does not prevent over fitting. For instance, consider a data set containing an even number of data points $N$ for which all $x$ and $y$ values are distinct. In this case one can split the observed $x$ values evenly between two $X$ bins while distributing one $y$ value into each of $N$ different $Y$ bins. This produces $MIC[x;y] = 1$, regardless of the actual $x$ and $y$ values in the data. The restriction to binning schemes satisfying $|X||Y| < B$ in Eq.\ \ref{MIC} circumvents this pathology; Reshef et al.\ advocate using either $B = N^{0.6}$ \cite{Reshef:2011p834} or $B = N^{0.55}$ \cite{Reshef:2013p1067}, but no mathematical rationale are given for these choices. 

\section*{$R^2$-based equitability is unsatisfiable}

Reshef et al.\ motivate MIC in large part by arguing that it satisfies their notion of equitability (described above), while existing measures like mutual information do not. However, no explicit mathematical definition of equitability was given in either the original \cite{Reshef:2011p834} or follow-up \cite{Reshef:2013p1067} work. For the purposes of this argument, we will adopt the following definition, which is consistent with the main text of the original paper \cite{Reshef:2011p834} as well as our discussions with the authors. To distinguish this definition from one presented in the next section, we will refer to this as ``$R^2$-equitability''.
 
\begin{definition}
A dependence measure $D[x;y]$ between two real stochastic variables $x$ and $y$ is \textbf{$R^2$-equitable} if and only if, in the large data limit,
\begin{eqnarray}
D[x;y] = g(R^2[f(x);y]), \label{bijection}
\end{eqnarray}
whenever 
\begin{eqnarray}
y = f(x) + \eta. \label{noisy_relationship}
\end{eqnarray}
Here, $f$ is a deterministic function of $x$, $\eta$ is a random noise term, $R^2[f(x);y]$ is the squared Pearson correlation between the noisy data $y$ and the noiseless value $f(x)$, and $g$ is an (unspecified) function that does not depend on $f$. Importantly, the only restriction we place on the noise $\eta$ is that it be drawn from a probability distribution which, if it depends on the value of $x$, does so only through the value of $f(x)$, i.e.\ $p(\eta|x) = p(\eta|f(x))$. A succinct way of stating this assumption, which we shall use repeatedly, is that the chain of variables $x \leftrightarrow f(x) \leftrightarrow \eta$ is a Markov chain.\footnote{Chapter 2 of \cite{Cover:1991p1068} provides additional information on Markov chains.}

\end{definition}

Heuristically this means that by computing $D[x;y]$ from knowledge of only $x$ and $y$, one can discern how tightly $y$ tracks the underlying noiseless value $f(x)$ without knowing what the function $f$ is. Reshef et al.\ claim that MIC satisfies $R^2$-equitability and that mutual information does not. However, they did not state what the function $g$ relating $MIC[x;y]$ to $R^2[f(x);y]$ is. And as mentioned above, no mathematical arguments were provided to support these claims.

It is readily shown, however, that MIC does not satisfy $R^2$-equitability: $MIC[x;y]$ is invariant to strictly monotonic transformations of $y$ and $f(x)$, but $R^2[f(x);y]$ is not, so no function $g$ can relate these two quantities.  

In fact, no nontrivial dependence measure $D[x;y]$ is $R^2$-equitable. To see this, choose the specific example of $y = x + \eta$ for arbitrary noise term $\eta$. Given any invertible function $h$, one can also write $y = h(x) + \mu$  where $\mu$ is a valid noise term.\footnote{Specifically, $\mu = h^{-1}(h(x)) - h(x) + \eta(h^{-1}(h(x)))$ is a stochastic variable who's distribution depends on $h(x)$ but not otherwise on $x$. Thus $x \leftrightarrow h(x) \leftrightarrow \mu$ is a Markov chain.} Thus,  $D[x;y] = g(R^2[x;y]) = g(R^2[h(x);y])$. But $R^2[x;y]$ is not invariant to invertible transformations of $x$. The function $g$ must therefore be constant, implying $D[x;y]$ cannot depend on the data and is therefore trivial.

We note that the supplemental material in \cite{Reshef:2011p834}, as well as the follow-up work \cite{Reshef:2013p1067}, suggest that $R^2$-equitability should be extended to include cases when noise is present in both the $x$ and $y$ directions. Formally, this means Eq.\ \ref{noisy_relationship} should read,
\begin{eqnarray}
y = f(x + \mu) + \eta \label{noisier_relationship}
\end{eqnarray}
for noise terms $\eta$ and $\mu$. But since no dependence measure satisfies $R^2$-equitability when $\mu = 0$, no measure can satisfy this stronger requirement. 

\section*{Self-consistent equitability}

It is therefore clear that the mathematical definition of equitability proposed by Reshef et al.\ cannot be adopted. The heuristic notion of equitability, however, remains valuable and is worth formalizing.  We therefore propose defining equitability instead as a self-consistency condition:

\begin{definition}
A dependence measure $D[x;y]$ is \textbf{self-equitable} if and only if 
\begin{eqnarray}
D[x;y] = D[f(x);y]\label{new_def}
\end{eqnarray} 
whenever $f$ is a deterministic function and $x \leftrightarrow f(x) \leftrightarrow y$ forms a Markov chain.
\end{definition} 

First note that this Markov chain condition includes relationships of the form shown in Eq.\ \ref{noisy_relationship}.\footnote{Adding noise in the $x$-direction through Eq.\ \ref{noisier_relationship}, however, violates this Markov chain requirement.} It also applies to other situations as well, e.g.\ where $x$ is a high dimensional vector, $y$ is categorical, and $f$ is a classifier function. Furthermore, self-equitability does not privilege a specific dependence measure such as the squared Pearson correlation $R^2$. Instead it simply asks for self-consistency: whatever value a dependence measure assigns to the relationship between $x$ and $y$, it must assign the same value to the dependence between $f(x)$ and $y$. 

Self-equitability is closely related to DPI, a critical inequality information theory that we now briefly review.
\begin{definition}
A dependence measure $D[x;y]$ satisfies the \textbf{Data Processing Inequality (DPI)} if and only if
\begin{eqnarray}
D[x;y] \leq D[z;y] \label{DPI}
\end{eqnarray}
whenever the stochastic variables $x,y,z$ form a Markov chain $x \leftrightarrow z \leftrightarrow y$.
\end{definition} 
DPI is an important requirement for any dependence measure, since it formalizes one's intuition that information is generally lost (and is never gained) when transmitted through a noisy communications channel. For instance, consider a game of telephone involving three children, and let the variables $x$, $z$, and $y$ represent the words spoken by the first, second, and third child respectively. The requirement stated in Eq.\ \ref{DPI} reflects our intuition that the words spoken by the third child will be more strongly dependent on those said by the second child (quantified by $D[z;y]$) than on those said by the first child (quantified by $D[x;y]$).

Every dependence measure satisfying DPI is self-equitable (see footnote\footnote{If $x \leftrightarrow z \leftrightarrow y$ is a Markov chain, then $f(x) \leftrightarrow x \leftrightarrow z \leftrightarrow y$ is a Markov chain as well for all deterministic functions $f$. If we further assume $z = f(x)$ for one specific choice of $f$, then $f(x) \leftrightarrow x \leftrightarrow f(x) \leftrightarrow y$ forms a Markov chain. DPI then implies $D[f(x);y] \leq D[x;y] \leq D[f(x);y]$, and so $D[x;y] = D[f(x);y]$. $\square$} for proof). In particular, mutual information satisfies DPI (see chapter 2 of \cite{Cover:1991p1068}) and is therefore self-equitable as well. Furthermore, any self-equitable dependence measure $D[x;y]$ must be invariant under \emph{all} invertible transformations of $x$ and $y$.\footnote{Given any joint distribution $p(x,y)$ and any two deterministic functions $h_1(x)$ and $h_2(y)$, it is straight-forward to show that $x \leftrightarrow h_1(x) \leftrightarrow y \leftrightarrow h_2(y)$ is a valid Markov chain. An equitable dependence measure $D$ must therefore satisfy $D[x;y] = D[h_1(x);y] = D[h_1(x),h_2(y)]$, proving the invariance of $D[x;y]$ under all invertible transformations of $x$ and $y$.  $\square$} MIC, although invariant under strictly monotonic transformations of $x$ and $y$, is not invariant under non-monotonic invertible transformations of either variable. MIC therefore violates both equitability and DPI. 

\begin{figure}[htbp] 
   \centering
   \includegraphics[width=4 in]{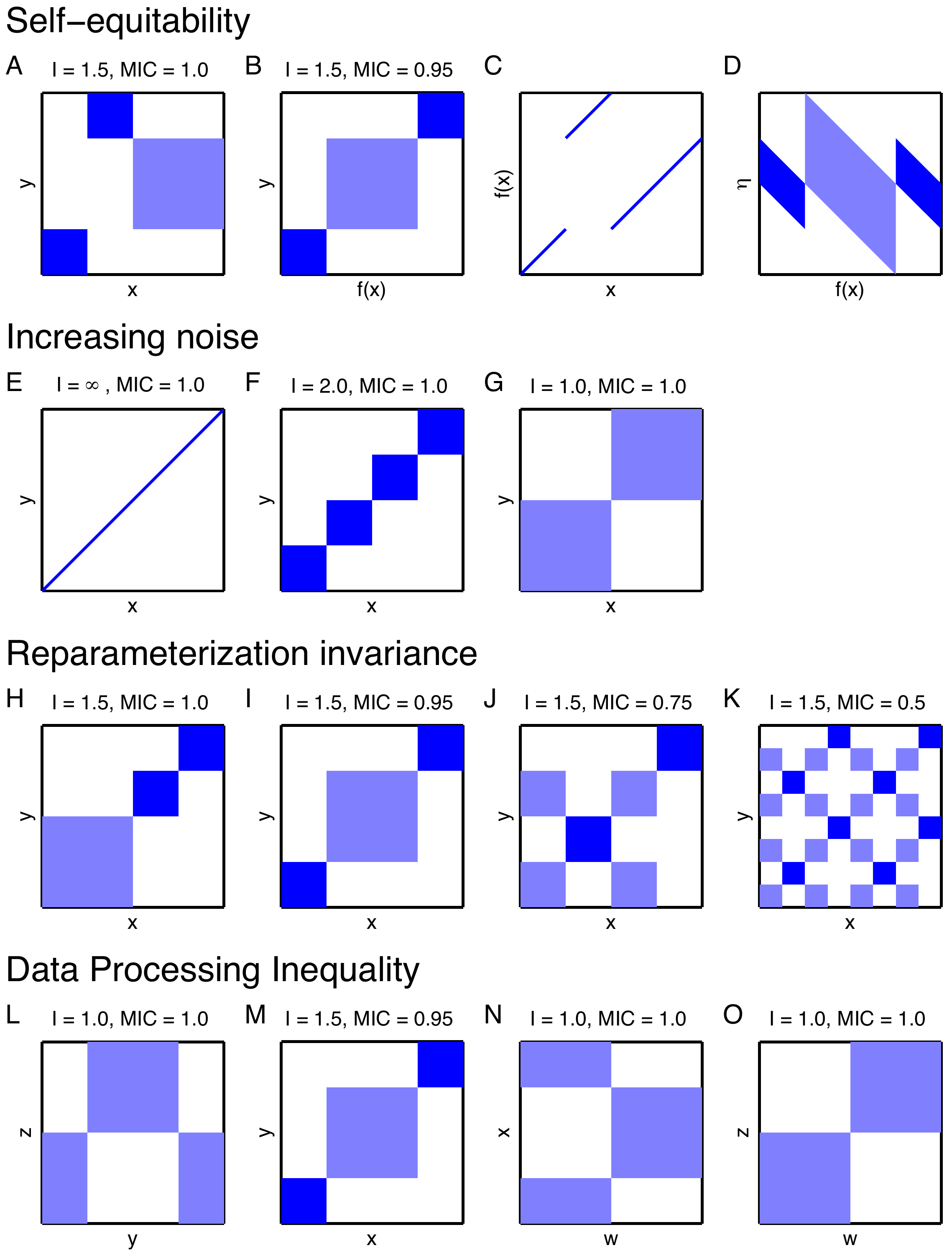} 
   \caption{MIC violates multiple notions of dependence that mutual information upholds. \textbf{(A,B,E-O)} Example relationships between two variables with indicated mutual information (denoted I) and MIC values.  Dark blue blocks represent twice the probability density as light blue blocks. \textbf{(A-D)} The relationships in (A,B) result from $y=f(x)+\eta$ with function $f(x)$ and noise term $\eta$ defined in (C,D). MIC is thus seen to violate self-equitability (Eq.\ \ref{new_def}) because $MIC[x;y] \neq MIC[f(x);y]$. Note that the noise term $\eta$, as required, depends on $f(x)$ but not otherwise on $x$. \textbf{(E-G)} Adding noise everywhere to the relationship in panel E diminishes mutual information but not  MIC. \textbf{(H-K)} The relationships in these panels are related by invertible non-monotonic transformations of $x$ and $y$; mutual information but not MIC is invariant under these transformations. \textbf{(L-O)} Convolving the relationships (panels L-N) linking variables in the Markov chain $w \leftrightarrow x \leftrightarrow y \leftrightarrow z$ produces the relationship between $w$ and $z$ shown in panel O. In this case MIC violates DPI because $MIC[w;z] > MIC[x;y]$; mutual information satisfies DPI here since $I[w;z] < I[x;y]$.}
   \label{fig:theory}
\end{figure}

\section*{Toy examples}

Fig.\ 1 demonstrates the contrasting behavior of mutual information and MIC in various simple relationships. Figs.\ 1A-D show an example relationship $y = f(x) + \eta$ for which MIC violates self-equitability. Figs.\ 1E-G show how adding noise everywhere reduces mutual information but does not always affect MIC; starting with a deterministic relationship $y=x$ (which has infinite mutual information), one can add noise to create various block diagonal relationships that have reduced mutual information, e.g.\ 2 bits (Fig.\ 1F) or 1 bit (Fig.\ 1G), but for which $MIC[x;y]=1$ remains saturated. Figs.\ 1H-K provide examples for which invertible non-monotonic transformations of $x$ and $y$ do not change $I[x;y]$ but greatly affect $MIC[x;y]$. Finally, Figs.\ 1L-O present a specific chain of relationships $w \leftrightarrow x \leftrightarrow y \leftrightarrow z$ illustrating how mutual information satisfies DPI  ($I[x;y] > I[w;z]$) while MIC does not ($MIC[x;y] < MIC[w;z]$).

\section*{Performance on simulated data}

We now revisit the simulation evidence offered by Reshef et al.\ in support of their claims about equitability. 

Reshef et al.\ first state that different noiseless relationships have different values for mutual information, and so mutual information violates their definition of equitability. To show this, they simulate 320 data points for a variety of deterministic relationships $y=f(x)$. They then estimate mutual information using the algorithm of Kraskov et al.\ \cite{Kraskov:2004p909} and observed that the values reported by this estimator vary depending on the function $f$ (Fig.\ 2A of \cite{Reshef:2011p834}). They state that this variation ``correctly reflect[s] properties of mutual information,'' and thus demonstrates that mutual information is not equitable.

\begin{figure}[htbp] 
   \centering
   \includegraphics[width=6in]{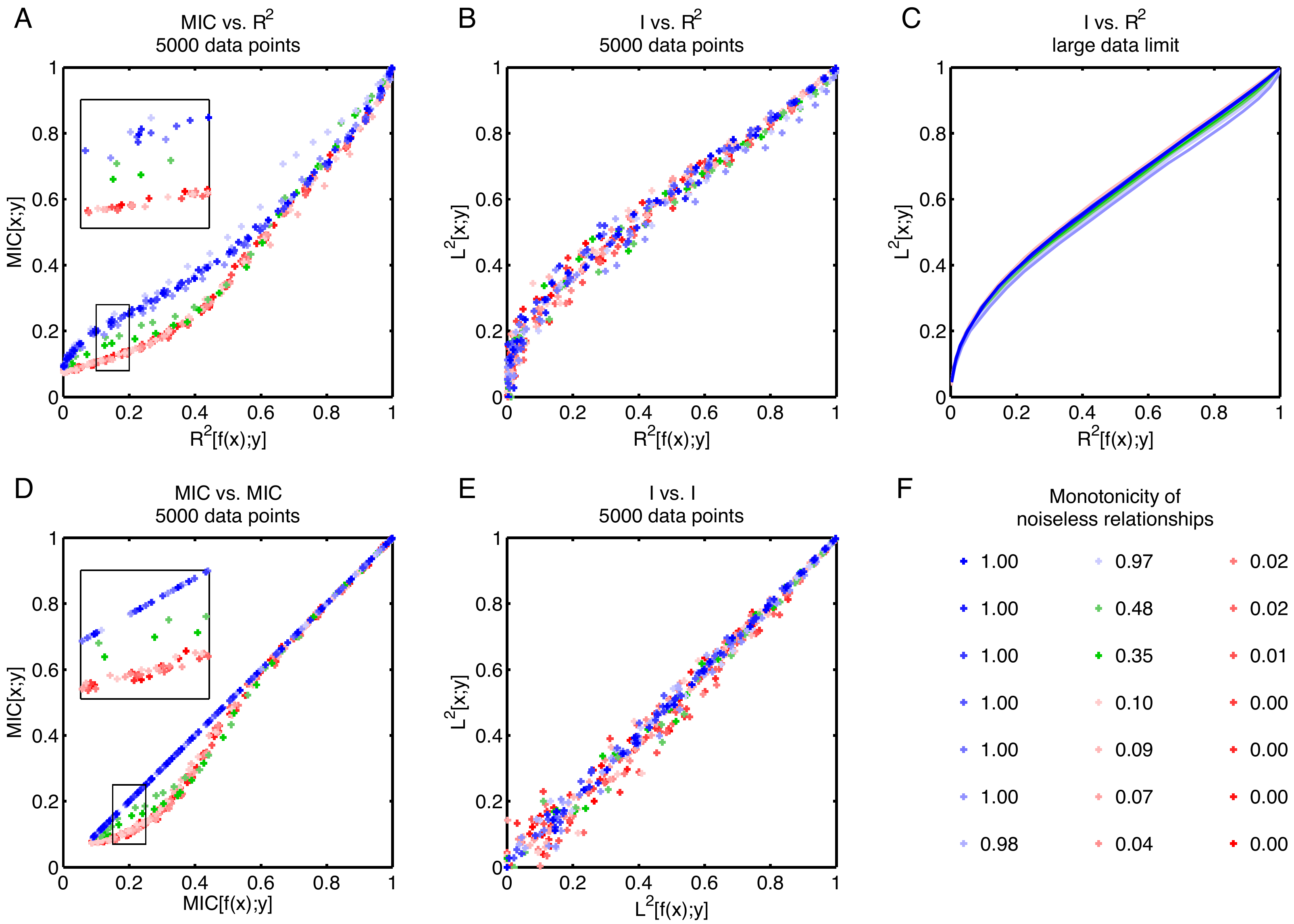} 
   \caption{Tests of $R^2$-equitability and self-equitability using simulated data. \textbf{(A,B,D,E)} Each plotted Ô+Õ represents results for 5000 data points generated as described for Figs.\ 2B,D of \cite{Reshef:2011p834}. MIC was computed using the algorithm of Reshef et al.\ with default settings. Mutual information was computed using the Kraskov et al.\ estimation algorithm \cite{Kraskov:2004p909} with smoothing parameter $k=1$. To facilitate comparison with MIC, mutual information values are represented in terms of the squared Linfoot correlation \cite{Speed:2011p910, Linfoot:1957p839}, which maps $I[x;y]$ (expressed in bits) to the interval $[0,1]$ via $L^2[x;y]=1 - 2^{-2I[x;y]}$. \textbf{(A)} $MIC[x;y]$ shows systematic dependence on the monotonicity of $f$ (see panel F) at fixed $R^2[f(x);y]$, thereby violating $R^2$-equitability. \textbf{(B,C)} $I[x;y]$ follows $R^2[f(x);y]$ much more closely than MIC, but slight deviations become evident in the large data limit. In panel C, mutual information was computed semianalytically using $I[x;y]=H[y] - H[\eta]$ where $H$ is entropy and $\eta$ is noise \cite{Cover:1991p1068, Rieke:1997p1060}. \textbf{(D,E)} MIC violates self-equitability because $MIC[x;y]$ deviates substantially from $MIC[f(x);y]$ depending on the monotonicity of $f$. This is not so for mutual information. \textbf{(F)} The monotonicity of each functional relationship, indicated by color, is quantified by the squared Spearman correlation between $x$ and $f(x)$. }
   \label{fig:simulations}
\end{figure}

It is clear that this observed variation results from finite sample effects, however, because deterministic relationships between continuous variables always have infinite mutual information.  Thus the different (finite) mutual information values reported can only result from imperfect performance of the specific mutual information estimator used. Indeed, this imperfect performance makes sense. The estimator of Kraskov et al.\ is optimized for data sets in which $k$ nearest neighbor data points in the $x$-$y$ plane (for some specified number $k$) are typically spaced much closer to one another than the length scale over which the joint distribution $p(x,y)$ varies significantly. Effectively, this estimator averages the joint distribution over $k$ nearest neighbor data points, with larger $k$ corresponding to greater smoothing. The default value of $k=6$, which was used by Reshef et al.\ (see \cite{Reshef:2013p1067}) is reasonable for many noisy real-world data sets, but may be inappropriate when the signal-to-noise ratio is high and data is sparse. In the limit of noiseless relationships, the primary assumption of the Kraskov et al.\ estimator is always violated.

For the case of noisy relationships, Reshef et al.\ simulated data of the form $y=f(x)+\eta$, generating between 250 and 1000 data points depending on $f$. They observed that, at fixed values of the squared Pearson correlation, $R^2[f(x);y]$, MIC, though varying substantially due to finite sample effects, did not exhibit a clear systematic dependence on the underlying function $f$. By contrast, the mutual information values returned by the Kraskov et al.\ estimator with $k=6$ showed a strong systematic dependence on $f$ (Figs.\ 2B,D of \cite{Reshef:2011p834}). However, we observed the opposite behavior when replicating this analysis using modestly larger data sets (having 5000 data points) and a minimal smoothing parameter ($k=1$) in the Kraskov et al.\ algorithm. Indeed, our estimates of $I[x;y]$ closely tracked $R^2[f(x);y]$ (Fig.\ 2B), and this behavior held approximately (but, as expected, not exactly) in the large data limit (Fig.\ 2C). By contrast, MIC depended strongly on the monotonicity of $f$ for $R^2 \lesssim 0.5$ (Fig.\ 2A,F). Thus, the relationship between $MIC[x;y]$ and $R^2[f(x);y]$ exhibits a clear dependence on the monotonicity of the underlying function $f$, even for the specific functions chosen by Reshef et al.. This agrees with the fact, shown above, that MIC does not satisfy $R^2$-equitability.\footnote{We note that Fig.\ 2 was shared with Reshef et al.\ prior to our posting this manuscript, and that panels A and B therein are reproduced in Fig.\ 7 of \cite{Reshef:2013p1067} (row 1, columns 1 and 2). However, Reshef et al.\ do not color data points according to the monotonicity of each underlying function as we do. This obscures the non-equitability of MIC. The dense overlay of points in their plots also exaggerates the noisiness of the mutual information estimates relative to the systematic bias observed in MIC.}

To test how well MIC obeys our definition of equitability (Eq.\ \ref{new_def}), we further compared $MIC[x;y]$ to $MIC[f(x);y]$ for these same simulated data sets, and again the relationship between these two values showed a clear dependence on the monotonicity of the function $f$ (Fig.\ 2D). Estimates of mutual information $I[x;y]$, however, traced estimates of $I[f(x);y]$ without apparent bias (Fig.\ 2E). 

Finally, to test the computational feasibility of estimating MIC as opposed to estimating mutual information, we timed how long it took for each estimator to produce the plots shown in Fig.\ 2. We observed the MIC algorithm of Reshef et al.\ to run $\sim600$ times slower than the Kraskov et al.\ mutual information estimation algorithm on these data, and this disparity increased dramatically with moderate increases in data size. Reshef et al.\ note that changing the runtime parameters of their algorithm can speed up MIC estimation \cite{Reshef:2013p1067}. This does not, however, appear to affect the poor scaling their algorithm's speed exhibits as $N$ increases (Table 3 of \cite{Reshef:2013p1067}). Thus, using the MIC estimation algorithm provided by Reshef et al.\ appears much less practical for use on large data sets than the mutual information estimation algorithm of Kraskov et al..
\section*{Discussion}

It should be emphasized that accurately estimating mutual information from sparse data is a nontrivial problem \cite{Treves:1995p244, Schurmann:2004p1054}. Indeed, many approaches for estimating mutual information (or entropy, a closely related quantity) have been proposed; a non-comprehensive list includes \cite{Slonim:2005p433, Kraskov:2004p909, Moddemeijer:1989p1050, Moon:1995p1049, Wolpert:1995p264, Nemenman:2001p1047, Nemenman:2004p1048, Schurmann:2002p1052, Chao:2003p1051, Paninski:2003p1046, Daub:2004p1065,  Cellucci:2005p1042, Hausser:2009p1043, Pal:2010p1039, Vinck:2012p1055}. While comparisons of various estimation methods have been performed \cite{Panzeri:2007p911, Khan:2007p1045, WaltersWilliams:2009p1066}, no single method has yet been accepted as decisively solving this problem in all situations. 

It has been argued \cite{Speed:2011p910} that the difficulty of estimating mutual information is one reason for using MIC as a dependence measure instead. However, MIC appears to be harder to estimate than mutual information both in principle and in practice. By definition, it requires one to explore all possible binning schemes for each data set analyzed. Consistent with this, we found the MIC estimator from \cite{Reshef:2011p834} to be much slower than the mutual information estimator of \cite{Kraskov:2004p909}. 

We are aware of two other critiques of the work by Reshef et al., one by Simon and Tibshirani \cite{Simon:2013p1024} and one by Gorfine et al.\ \cite{Gorfine:2012p1028}. These do not address the issues of equitability discussed above, but rather focus on the statistical power of MIC. Through the analysis of simulated data, both critiques found MIC to be less powerful than a recently developed statistic called ``distance correlation'' (dCor) \cite{Szekely:2009p1034}. Gorfine et al.\ \cite{Gorfine:2012p1028} also recommend a different statistic, HHG \cite{Heller:2012p1064}. Like mutual information, both dCor and HHG can detect arbitrary relationships between vector-valued variables. Moreover, both dCor and HHG are ``plug-in'' statistics that can be easily computed directly from data, i.e.\ they do not require an approximate estimation procedure like mutual information and MIC do. However, neither $dCor[x;y]$ or $HHG[x;y]$ are invariant under invertible transformations of $x$ and $y$. As a result, both statistics violate self-equitability as well as DPI. We therefore suggest that dCor or HHG may be useful in situations when estimating mutual information is either impractical, due to computational cost or under-sampling, or does not provide sufficient statistical power.

Certain mutual information estimators, however, do work well enough to be used in many real-world situations.\footnote{In particular, we found the estimator of Kraskov et al.\ to perform admirably on the simulated data generated for Fig.\ 2.} This is evidenced by the fact that mutual information has been used to tackle a variety of problems in many fields including neuroscience \cite{Rieke:1997p1060, Sharpee:2004p319, Sharpee:2006p571}, molecular biology \cite{Kinney:2007p89, Kinney:2010p293, Elemento:2007p119, Goodarzi:2012p943}, medical imaging \cite{Pluim:2003p1029}, and signal processing \cite{Hyvarinen:2000p1037}. We also note that mutual information has been used by multiple groups to construct networks analogous to those constructed by Reshef et al.\ using MIC \cite{Slonim:2005p433, Butte:2000p1026, Steuer:2002p1025,  Margolin:2006p435, Shi:2011p1040}. 

We emphasize, however, that \emph{all} the difficulties associated with estimating mutual information vanish in the large data limit. Mutual information estimation appears difficult in the analysis of Reshef et al.\ primarily because of their focus on relatively small high-dimensional data sets.  Indeed, none of the simulations in \cite{Reshef:2011p834} comprise more than 1000 data points. While small high-dimensional data sets do often occur, well-sampled data sets are becoming increasingly prevalent. Consumer research companies routinely analyze data sets containing information on $\sim 10^5$ shoppers, while companies like Facebook and Google can access information on $\sim 10^9$ people. Banks and hedge funds routinely comb through databases containing years of stock and commodity prices recorded at sub-second resolution. In biology, DNA sequencing technology is driving massive data generation. For instance, a relatively small scale experiment by Kinney et al.\ \cite{Kinney:2010p293} used DNA sequencing to measure the transcriptional activities of $2.5 \times 10^5$ mutants of a specific transcriptional regulatory sequence only 75 nucleotides in length.\footnote{Mutual information played a critical role as a dependency measure in the analysis of these data.} Among larger scale scientific efforts, the U.K. recently announced plans to sequence the genomes of $10^5$ people -- much more than the number of genes in the human genome.  

Mutual information is an important tool for making sense of such well-sampled data sets. Not only does it naturally quantify the strength of relationships between arbitrary variables, its close connection to likelihood  \cite{Kinney:2012p1023, Kouh:2009p939} makes it a proper objective function for fitting parameterized models to a wide variety of data sets \cite{Sharpee:2006p571, Kinney:2007p89, Kinney:2010p293, Elemento:2007p119, Goodarzi:2012p943}. Although the mutual information estimation problem has not been solved definitively, it has been solved well enough for many practical purposes, and all lingering difficulties vanish in the large data limit. We therefore believe that mutual information has the potential to become a critical tool for making sense of the large data sets proliferating across disciplines, both in science and in industry. 

\section*{Acknowldegements}

Acknowledgments: We thank Bruce Stillman, Bud Mishra, and Swagatam Mukhopadhyay for their useful feedback. This work was supported by the Simons Center for Quantitative Biology at Cold Spring Harbor Laboratory (J.B.K. and G.S.A.) and STARR Cancer Consortium grant 13-A123 (G.S.A.). The authors declare no conflicts of interest.

\bibliography{polished_library}{}
\bibliographystyle{ieeetr}
\end{document}